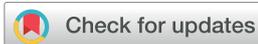

# S66x8 noncovalent interactions revisited: new benchmark and performance of composite localized coupled-cluster methods†

Golokesh Santra, ‡[a] Emmanouil Semidalas, ‡[a] Nisha Mehta, [a] Amir Karton *[bc] and Jan M. L. Martin *[a]

The S66x8 noncovalent interactions benchmark has been re-evaluated at the "sterling silver" level, using explicitly correlated MP2-F12 near the complete basis set limit, CCSD(F12*)/aug-cc-pVTZ-F12, and a (T) correction from canonical CCSD(T)/sano-V{D,T}Z+ calculations. The revised reference values differ by 0.1 kcal mol⁻¹ RMS from the original Hobza benchmark and its revision by Brauer *et al.*, but by only 0.04 kcal mol⁻¹ RMS from the "bronze" level data in Kesharwani *et al.*, *Aust. J. Chem.*, 2018, **71**, 238–248. We then used these to assess the performance of localized-orbital coupled cluster approaches with and without counterpoise corrections, such as PNO-LCCSD(T) as implemented in MOLPRO, DLPNO-CCSD(T₁) as implemented in ORCA, and LNO-CCSD(T) as implemented in MRCC, for their respective "Normal", "Tight", and "very Tight" settings. We also considered composite approaches combining different basis sets and cutoffs. Furthermore, in order to isolate basis set convergence from domain truncation error, for the aug-cc-pVTZ basis set we compared PNO, DLPNO, and LNO approaches with canonical CCSD(T). We conclude that LNO-CCSD(T) with veryTight criteria performs very well for "raw" (CP-uncorrected), but struggles to reproduce counterpoise-corrected numbers even for veryveryTight criteria: this means that accurate results can be obtained using either extrapolation from basis sets large enough to quench basis set superposition error (BSSE) such as aug-cc-pV{Q,5}Z, or using a composite scheme such as Tight{T,Q} + 1.11[vvTight(T) − Tight(T)]. In contrast, PNO-LCCSD(T) works best with counterpoise, while performance with and without counterpoise is comparable for DLPNO-CCSD(T₁). Among more economical methods, the highest accuracies are seen for dRPA75-D3BJ, ωB97M-V, ωB97M(2), revDSD-PBEP86-D4, and DFT(SAPT) with a TDEXX or ATDEXX kernel.

## I. Introduction

Noncovalent interactions (NCIs) are crucial to several chemical and biological phenomena occurring in the solid, liquid, and gaseous phases.[1–3] These interactions play a pivotal role in designing new functional materials,[4–7] controlling solvation dynamics,[8–10] protein folding,[11] and catalysis[12,13] with applications in, *e.g.*, liquid-crystal technology,[14] drug delivery,[15] and many more. Over the decades, several experimental studies have been performed (see ref. 1, 16, 17 and references therein).

NCIs in the condensed phase are often determined *via* NMR spectroscopy,[18–20] while in the gas phase, rotational spectroscopy is a reliable technique for polar molecules; for recent studies, see ref. 21–24. Significant efforts are still ongoing to understand and quantitatively measure different types of noncovalent interactions, present in bulk as well as interfaces. Unfortunately, results from experimentally studied noncovalently-bonded systems often include dynamical and (or) environmental effects, thus rendering them inappropriate to use as a reference for developing semi-empirical methods. Hence, accurate wavefunction *ab initio* methods are often essential for obtaining highly precise NCI energies. Over the past few years, several datasets have been proposed[25–38] for intra- and intermolecular noncovalent interactions involving biologically important organic molecules of different sizes. The strong and weak NCIs in biomolecules are particularly interesting because they often play prominent roles in determining their favorable structures, specific binding sites, and dynamics.[11] Hobza's S66[35] and its extended version, S66x8,[37] are two such datasets. The S66x8 set comprises equilibrium and angular-displaced nonequilibrium geometries of 66 dimers—built from 14 different monomers representative of biomolecule fragments. These nonequilibrium structures were generated by multiplying the equilibrium intermonomer distances ($r_e$) by factors of 0.90, 0.95, 1.00, 1.05, 1.10, 1.25, 1.50, and 2.00, respectively, while freezing the other degrees of freedom. All dimers of S66x8 can be categorized into four subsets: hydrogen bonds, π-stacking, London dispersion complexes, and mixed-influence interactions.

*[a] Department of Molecular Chemistry and Materials Science, Weizmann Institute of Science, 7610001 Rehovot, Israel. E-mail: gershom@weizmann.ac.il*
*[b] School of Molecular Sciences, The University of Western Australia, Perth, WA 6009, Australia*
*[c] School of Science and Technology, University of New England, Armidale, NSW 2351, Australia*
† Electronic supplementary information (ESI) available. See DOI: https://doi.org/10.1039/d2cp03938a
‡ Shared first authors.

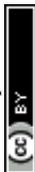

It is already well established that even at the complete basis set (CBS) limit, second-order Møller–Plesset perturbation theory (MP2) performs poorly for atomization energies, electron affinities, and barrier heights (see ref. 39 and references therein). However, for noncovalent interactions, MP2 with a large basis set can be considered a starting point, which, combined with high-level corrections (e.g., CCSD(T)–MP2 in a smaller basis), can lead to more accurate ab initio energies.[35,37] [CCSD(T) refers to coupled-cluster theory, including singles, doubles, and perturbative triples[40,41].]

The original S66x8 NCI reference energies were obtained by adding a [CCSD(T)-MP2]/AVDZ high-level correction (HLC) to the extrapolated MP2/AV{T,Q}Z with full counterpoise (CP) correction. In ref. 42 Brauer et al. improved the reference energies by combining explicitly correlated MP2-F12 energies at the CBS limit with an HLC from [CCSD(T$_{sc}$)-F12b − MP2-F12]/cc-pVDZ-F12. Depending on the ab initio methods used for high-level correction, Martin and coworkers[43] proposed a hierarchy of revisited standards for NCI energies: gold, silver or sterling silver, and bronze (see Section IIIa for further details). In practice, the "sterling silver" (i.e., an alloy of 92.5% pure silver and 7.5% other metals, usually copper) level NCI energies are a low-cost version of the original "silver" standard. Due to the daunting computational cost, the authors in ref. 43 computed the interaction energies at the "gold" level for only 18 dimers out of 66 present in S66. However, they successfully calculated the "sterling silver" and "bronze" standard energetics of the S66 and S66x8 sets, respectively.

Although canonical CCSD(T) or explicitly correlated CCSD(T)-F12 is often desired for highly accurate NCI energies, the steep $N^7$ cost scaling of these methods with the system size ($N$) have made them prohibitively expensive for applications on larger systems. Hence, over the past few years, localized coupled-cluster methods, such as the PNO-LCCSD(T) [pair natural orbital localized coupled-cluster with singles, doubles, and perturbative triples] approach of the Werner group,[44] DLPNO-CCSD(T) [domain localized pair natural orbital CCSD(T)] and related methods by the Neese group,[45,46] and the LNO-CCSD(T) [localized natural orbital CCSD(T)] method of Kallay and coworkers[47–49] have gained considerable attention. These methods' attractive linear scaling behavior (for sufficiently large systems) allows calculations on systems consisting of hundreds of atoms. Often, with tight accuracy cutoffs and large basis sets, they can achieve accuracy comparable to canonical CCSD(T). Examples of recent use include the energetics of the $(H_2O)_{20}$ cages using PNO-LCCSD(T)-F12b,[50]

(the F12b suffix refers to explicit correlation[51]), the noncovalent interaction energies of seven large dimers (L7 set[32]) with LNO-CCSD(T),[52] the main group thermochemistry, barrier heights, intra-, and intermolecular interaction energies of GMTKN55[53] using DLPNO-CCSD(T),[54] and benchmark studies on the Ru(II) complexes involved in hydroarylation,[55] highly delocalized polypyrroles (POLYPYR[56] set), and metal–organic barrier heights (MOBH35, 35 reactions[57–59]).

Granted the linear cost scaling, the accuracy of localized approaches is subject to various predefined cutoffs, and tuned fixed combinations of these cutoffs are given as keywords in several codes. The available options for DLPNO-CCSD(T) and related methods in ORCA[60] are LoosePNO, NormalPNO, TightPNO, and VeryTightPNO (see Table 1 in ref. 61 for definitions). The accuracy presets in LNO-CCSD(T) are set to Normal, Tight, vTight, or vvTight (see Table 1 in ref. 49 for details). In PNO-LCCSD(T), Default and Tight (see Tables 1–4 in ref. 44 for more information) are the standard presets.

Recently, S66 noncovalent interactions were studied with LNO-CCSD(T) by Kállay and Nagy,[49] and with eight low-cost LNO-CCSD(T)-based composite schemes by ourselves, succinctly reported in a conference proceeding extended abstract.[62] Both studies showed a smooth error convergence by gradually tightening the accuracy thresholds and increasing the basis set size with respect to the "silver"[43] standard reference energies. For our composite schemes, a two-point CBS extrapolation with the same accuracy cutoff was employed, and the effect of a tighter cutoff was estimated as a scaled additivity correction in a relatively small basis set. With half-counterpoise[63,64] correction (i.e., the average of full counterpoise-corrected and uncorrected results), some low-cost composite LNO-CCSD(T) methods performed very well.

That being said, the main objectives of the present study can be briefly summarized as: (a) climbing one step of the hierarchical ladder of NCI reference energies toward "sterling silver" standard S66x8 interaction energies; (b) evaluating the performance of pure and composite localized coupled-cluster methods based on LNO-CCSD(T), PNO-LCCSD(T), and DLPNO-CCSD(T$_1$) against the S66x8 set; and (c) recommending a localized coupled cluster method which can be used for calculating accurate NCI energies, thus avoiding the expensive canonical coupled-cluster options.

## II. Computational methods

The Molpro 2021.3[65] program suite was used for the conventional and explicitly correlated ab initio single-point calculations. The MRCC 2020,[66] Molpro 2021.3,[65] and ORCA 5.0.1[60] packages were employed for the LNO-CCSD(T), PNO-CCSD(T), and DLPNO-CCSD(T$_1$) calculations, respectively. The reference geometries of the S66x8 dataset were extracted from the Benchmark Energy & Geometry Database website (https://www.begdb.org/) and were used without further optimization. All electronic structure calculations were performed jointly on the Faculty of Chemistry's Linux cluster, ChemFarm, at the Weizmann

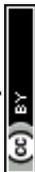

Institute of Science, and the Karton group's Linux cluster at the University of Western Australia (UWA).

For the explicitly correlated RI-MP2-F12 with the familiar 3C(FIX) ansatz,[67,68] we used the correlation-consistent cc-pV$n$Z-F12 basis sets[69] for the atom of hydrogen and aug-cc-pV$n$Z-F12[70] ($n$ = T, Q) for the nonhydrogen atoms; this expedient is denoted haV$n$Z-F12 ($n$ = T, Q) in the present manuscript. The Boys and Bernardi counterpoise correction (CP)[71] was also used for removing the ''basis set superposition errors'' (BSSE).[72,73] For the CCSD(F12*)[74,75] (known in Molpro-speak as CCSD-F12c), we used the non-augmented cc-pVTZ-F12[69] throughout.

For the MP2-F12 steps, the 3C(D)[76] approach might have led to somewhat faster basis set convergence; however, while it is both size-consistent and geminal BSSE-free like 3C(Fix), it is not orbital-invariant. As we were already using quite large basis sets, we opted for the orbital-invariant 3C(Fix) default instead, which use fixed geminal amplitudes fixed by the cusp conditions.[77]

CCSD-F12a,[78] CCSD-F12b,[78] and CCSD(F12*),[74,75] represent progressively more rigorous approximations to exact CCSD-F12—for a detailed analysis see Köhn and Tew[75] and ref. 79—the latter of which has extensive numerical data for total atomization energies. For interaction energies of water clusters, Manna et al.[80] showed clearly that CCSD(F12*) exhibits the fastest basis set convergence of all three methods. Specifically for S66, the various approximations to CCSD-F12 were investigated in some detail,[43] and CCSD(F12*) was found to be the most accurate by far. The repeatedly made claim, e.g. ref. 81, that CCSD-F12a is superior to CCSD-F12b for small basis sets, rests on an error compensation between basis set incompleteness of these small basis sets and the known overbinding[43,79,80] of F12a. The senior authors eschew ''right answers for the wrong reasons'' to the greatest extent possible.

For the conventional CCSD and CCSD(T), we employed the semi-augmented sano-pV$n$Z+ ($n$ = D, T) atomic natural orbital basis sets of Neese and Valeev.[82] Appropriate auxiliary basis sets for JKfit[83] (Coulomb and exchange) and MP2fit[84,85] (density fitting in MP2) were used across the board. As recommended by Werner and coworkers[86] for VTZ-F12 and VQZ-F12, the geminal exponent $\beta = 1.0a_0^{-1}$ was employed for the haVTZ-F12 and haVQZ-F12 basis sets.

For the localized CCSD(T) calculations, haV$n$Z ($n$ = T, Q, and 5) basis sets, together with the corresponding haV$n$Z/MP2fit[84,85] ($n$ = T, Q, and 5), were used. As the present paper was being revised, ref. 87 was brought to our attention, where it was found that substituting haV($n$+1)Z/MP2Fit for haV$n$Z/MP2Fit significantly reduces error in the absolute correlation energy. We tested this at 1.0$r_e$ for water dimer (system 1) and benzene dimer pi-stacked (system 24) with Tight cutoffs and the haVTZ and haVQZ basis sets. While the changes in absolute correlation energies are close to what is reported in the ESI of ref. 87, they cancel almost exactly between dimer and separate monomers, such that the interaction energies remain unchanged to two decimal places (see also discussion of Tables 4 and 5 in Sherrill et al.[88]).

Concerning accuracy thresholds, Normal, Tight, vTight, and vvTight were considered for LNO-CCSD(T), whereas for the

PNO-LCCSD(T) calculations, we applied the Default and Tight settings. On the other hand, for DLPNO-CCSD(T$_1$), we employed NormalPNO, TightPNO, and VeryTightPNO (TightPNO with $T_{CutMKN}$, $T_{CutPNO}$, and $T_{CutPairs}$ tightened to $10^{-4}$, $10^{-8}$, and $10^{-6}$, respectively.[89] For the detailed description of these cutoff parameters, see ref. 61) settings together with the RIJCOSX (resolution of the identity in combination with the chain of spheres[90] algorithm) approximation. (T$_1$) refers to the more rigorous full iterative triples correction,[91] and (T$_0$), often confusingly referred to as (T), refers to the older, more economical noniterative approach[46] in which the off-diagonal Fock matrix elements are neglected. Unlike (T$_0$), (T$_1$) entails nontrivial I/O bandwidth requirements.

To investigate the dependence of the DLPNO-CCSD(T$_1$) correlation on the size of the PNO space, we also adjusted the $T_{CutPNO}$ threshold in DLPNO-CCSD(T$_1$)/Tight from the default value $10^{-7}$ to $10^{-6}$. Two-point PNO extrapolations were also carried out to the complete PNOs space limit (CPS), using the simple two points extrapolation scheme proposed by Altun et al.,[92] $E = E^X + [Y^\beta/(Y^\beta - X^\beta)] \times (E^Y - E^X)$, where $Y = X + 1$ and $\beta = 7.13$. This corresponds numerically to $E^X + 1.5 \times (E^Y - E^X)$.

Similar to our previous studies,[62,64,80] in addition to full counterpoise (CP) corrections,[71] we have also considered the average of raw and CP, i.e., 'half-CP', to calculate the dissociation energies of S66x8. A two-point basis set extrapolation was carried out using the following expression from Halkier et al.,[93]

$$E_{CBS} = E_L + (E_L - E_{L-1}) \Big/ \left[ \left( \frac{L}{L-1} \right)^\alpha - 1 \right],$$ where $L$ refers to the basis set cardinal number and $\alpha$ is the basis set extrapolation exponent. For RI-MP2-F12/haV{T,Q}Z-F12, where haV{T,Q}Z-F12 denotes the extrapolation from haVTZ-F12 and haVQZ-F12 basis sets, we used the same $\alpha$ value of 4.6324 initially obtained for the S66 set.[43] Like the W1 and W2 theories,[94] for the localized CC methods, we employed $\alpha = 3$ and 3.22 for the haV{Q,5}Z and haV{T,Q}Z extrapolations, respectively. The optimal values of the linear combination coefficients for the composite localized CC schemes were obtained by minimizing the root mean square deviations (RMSD) from the new reference energies.

Most of the DFT results were extracted from the ESI of ref. 42; the revDSD functionals from ref. 95 were evaluated using ORCA, while for $\omega$B97M-V[96] and $\omega$B97M(2),[97] the reference implementations in Q-CHEM 5 were used.[98]

## III. Results and discussion

### (a) Re-evaluation of reference energies

We were able to calculate the RI-MP2-F12/haVTZ-F12 and RI-MP2-F12/haVQZ-F12 level dissociation energies of the S66x8 dimers with and without counterpoise (CP) correction. With respect to the full-CP corrected RI-MP2-F12/haV{T,Q}Z-F12 energies, the RMS error of the counterpoise-uncorrected (i.e., ''raw'') and half-CP are 0.005 and 0.002 kcal mol$^{-1}$, respectively. Hence, either full-CP, half-CP, or ''raw'' MP2-F12/CBS energies can be a starting point for re-evaluating the S66x8 reference NCI energies.

Originally proposed[43] for the S66, the four types of HLCs are: (a) *gold* (employing [CCSD(F12*)–MP2-F12]/cc-pVQZ-F12 half-CP combined with (T)/haV{T,Q}Z half-CP), (b) *silver* (using [CCSD(F12*)–MP2-F12]/cc-pVTZ-F12 half-CP plus (T)/haV{D,T}Z half-CP), (c) *sterling silver* (combining [CCSD(F12*)–MP2-F12]/cc-pVTZ-F12 raw and (T)/sano-PV{D,T}Z+ raw), and (d) *bronze* (employing half-CP CCSD(F12*)(Tsc)/cc-pVDZ-F12). In the present study, we adopt the "sterling silver"-level HLC on top of the full-CP corrected RI-MP2-F12/haV{T,Q}Z-F12 noncovalent interaction energies, *i.e.*, RI-MP2-F12/CBS is combined with counterpoise uncorrected [CCSD(F12*)–MP2-F12]/cc-pVTZ-F12, and [CCSD(T)–CCSD]/sano-pV{D,T}Z+. The equation for the final energy looks like this:

$$E_{Ref} = E_{MP2-F12/haV\{T,Q\}Z-F12}^{CP} + E_{[CCSD(F12*)-MP2-F12]/VTZ-F12}^{RAW} + E_{[CCSD(T)-CCSD]/sano-pV\{D,T\}Z+}^{RAW}$$

Our best estimates of the S66x8 dissociation energies are listed in Table 1. Relative to the "sterling silver" standard dissociation energies, the RMS deviation for Hobza's[37] original S66x8 reference is 0.103 kcal mol⁻¹, which is marginally reduced to 0.096 kcal mol⁻¹ for the revised reference proposed by Brauer *et al.*[42] However, the "bronze" quality dissociation energies lately proposed by Kesharwani *et al.*[43] have only 0.041 kcal mol⁻¹ deviation. Among the four subsets, the most significant contribution to the RMS error comes from the π-stack complexes. In general, the error is most prominent for the compressed geometries, which gets dramatically reduced for the relaxed cases (see Table 2).

As the present paper was being finalized for submission, a paper by Nagy, Kállay, and coworkers[99] was published in which they revisited the S66 reference energies at what they call a "14-karat" [sic] "gold" (14k-gold) level. Their levels of theory through CCSD are nearly identical to those used in the "silver" and "sterling silver" datasets of ref. 43, but their (T) treatment is effectively that of the "gold" level (used only for a subset of 18 systems ref. 43 owing to computational cost), namely CCSD(T)/haV{T,Q}Z albeit with a DF approximation.

Essentially all of the discrepancies between "14k-gold" on the one hand and "silver" and "sterling silver" on the other hand can be attributed to the more economical triples corrections in these latter reference standards, namely CCSD(T)/haV{D,T}Z and CCSD(T)/sano-{D,T}Z+, respectively.

For the 18 systems where we were able to obtain "gold" answers in ref. 43, the 14k-gold values of Nagy *et al.*[99] agree to 0.008 kcal mol⁻¹ RMS. The "silver" is very close to both, with RMSD of just 0.011 kcal mol⁻¹ from the "14k-gold" for the whole S66 set, and of just 0.006 kcal mol⁻¹ from the "gold" for the subset of 18 systems where the latter is available. In contrast, the more economical "sterling silver" level used for S66x8 in the present study deviates by a somewhat larger 0.027 kcal mol⁻¹ from "14k-gold" for S66. For the 18-system "gold" subset, we find a not dissimilar 0.024 kcal mol⁻¹ deviation.

Keeping also in mind residual uncertainties in other components, we conclude from the above that 0.03 kcal mol⁻¹ RMS is a realistic estimate of the uncertainty in our revised S66x8 data.

## (b) LNO-CCSD(T)-based methods

In ref. 62 we benchmarked the performance of pure and composite LNO-CCSD(T) methods with respect to the S66 noncovalent interaction energies ("silver" standard[43]). In the present study, we investigate the performance of LNO-CCSD(T)-based methods for the S66x8 set using the different basis sets and accuracy thresholds available.

Like in ref. 62 here too, we find a consistent improvement of accuracy with increasing basis set size and tightening of the thresholds; counterpoise correction also offers an appreciable improvement (see upper blocks of Table 3). With vTight threshold and half-CP, LNO-CCSD(T)/haV5Z is the best pick (0.056 kcal mol⁻¹) among the single basis set approaches tested here. Except for the "Normal" settings, the RMS deviations with half-CP are marginally better than the full-CP option for all other accuracy thresholds.

Irrespective of the choice of accuracy threshold, a two-point CBS extrapolation improves the RMS error statistics across the board. Among all the LNO-CCSD(T)/CBS methods listed in Table 3, the LNO-CCSD(T, vTight)/haV{T,Q}Z half-CP is the best pick (0.025 kcal mol⁻¹) for S66x8 noncovalent interactions. With the "Normal" setting and counterpoise uncorrected interaction energies, the low-cost LNO-CCSD(T)/haV{T,Q}Z fortuitously outperforms LNO-CCSD(T)/haV{Q,5}Z, and gradually tightening the accuracy threshold narrows that performance gap. The LNO-CCSD(T)/CBS methods with half-CP have slightly lower RMSD than their full-CP counterparts; however, the differences are within the uncertainty of the reference data. (Needless to say, at the true CBS limit, the difference between raw and CP-corrected values should be zero: a significant discrepancy between raw and CP in a CBS extrapolation suggests that the underlying calculations are inadequate, or that the extrapolation procedure is problematic.)

Closer scrutiny of LNO-CCSD(T) performance for the subsets of S66x8 reveals that, for any individual haVnZ basis set, CP correction always degrades performance for hydrogen bonds, but improves it for the three other subsets. Upon CBS extrapolation, we find a significant raw-CP difference for hydrogen bonds with Normal cutoffs and haV{Q,5}Z basis sets, favoring full CP, while the opposite is true for the other three subsets. With Tight cutoffs, the raw-CP differences upon extrapolation approach the uncertainty in the reference values; with vTight cutoffs, one can no longer meaningfully choose between raw and CP. One might elect to use half-CP and use the difference between raw and CP as a crude uncertainty bar; on the other hand, the "raw" calculations are more convenient, especially when applied to intramolecular interactions where CP correction would be impractical. Using the vTight threshold and half-CP, LNO-CCSD(T)/haV{T,Q}Z and LNO-CCSD(T)/haV5Z are statistically tied for best pick for all four subsets of S66x8 (see the

**Table 1** Sterling silver[a] level S66x8 dissociation energies (kcal mol$^{-1}$). $r_e$ represents the equilibrium distance for each complex

| | Dimers | 0.90$r_e$ | 0.95$r_e$ | 1.00$r_e$ | 1.05$r_e$ | 1.10$r_e$ | 1.25$r_e$ | 1.50$r_e$ | 2.00$r_e$ |
|---|---|---|---|---|---|---|---|---|---|
| **Hydrogen bonding** | | | | | | | | | |
| 1 | Water···Water | 4.666 | 4.954 | 4.946 | 4.762 | 4.480 | 3.470 | 2.117 | 0.874 |
| 2 | Water···MeOH | 5.298 | 5.629 | 5.627 | 5.423 | 5.106 | 3.956 | 2.391 | 0.954 |
| 3 | Water···MeNH$_2$ | 6.606 | 6.957 | 6.941 | 6.695 | 6.318 | 4.928 | 2.980 | 1.142 |
| 4 | Water···Peptide | 7.739 | 8.131 | 8.128 | 7.878 | 7.482 | 5.994 | 3.830 | 1.439 |
| 5 | MeOH···MeOH | 5.387 | 5.771 | 5.804 | 5.620 | 5.312 | 4.151 | 2.528 | 1.011 |
| 6 | MeOH···MeNH$_2$ | 7.081 | 7.554 | 7.601 | 7.378 | 6.994 | 5.505 | 3.345 | 1.273 |
| 7 | MeOH···Peptide | 7.756 | 8.235 | 8.289 | 8.072 | 7.692 | 6.188 | 3.647 | 1.098 |
| 8 | MeOH···Water | 4.675 | 5.017 | 5.045 | 4.881 | 4.610 | 3.598 | 2.205 | 0.909 |
| 9 | MeNH$_2$···MeOH | 2.837 | 3.051 | 3.037 | 2.892 | 2.682 | 1.971 | 1.098 | 0.393 |
| 10 | MeNH$_2$···MeNH$_2$ | 3.722 | 4.104 | 4.153 | 4.002 | 3.742 | 2.781 | 1.302 | 0.385 |
| 11 | MeNH$_2$···Peptide | 4.958 | 5.354 | 5.397 | 5.222 | 4.923 | 3.197 | 1.397 | 0.455 |
| 12 | MeNH$_2$···Water | 6.861 | 7.291 | 7.312 | 7.077 | 6.691 | 5.231 | 3.155 | 1.196 |
| 13 | Peptide···MeOH | 5.753 | 6.165 | 6.220 | 6.054 | 5.759 | 4.608 | 2.948 | 1.307 |
| 14 | Peptide···MeNH$_2$ | 6.907 | 7.411 | 7.500 | 7.324 | 6.986 | 5.610 | 3.549 | 1.488 |
| 15 | Peptide···Peptide | 8.096 | 8.586 | 8.659 | 8.464 | 8.105 | 6.649 | 4.407 | 1.775 |
| 16 | Peptide···Water | 4.761 | 5.108 | 5.150 | 5.005 | 4.754 | 3.799 | 2.454 | 1.137 |
| 17 | Uracil···Uracil (BP) | 16.066 | 17.187 | 17.405 | 17.034 | 16.293 | 13.190 | 8.357 | 3.337 |
| 18 | Water···Pyridine | 6.484 | 6.870 | 6.879 | 6.653 | 6.289 | 4.929 | 3.006 | 1.184 |
| 19 | MeOH···Pyridine | 6.876 | 7.374 | 7.448 | 7.253 | 6.897 | 5.473 | 3.373 | 1.328 |
| 20 | AcOH···AcOH | 17.925 | 19.170 | 19.397 | 18.965 | 18.125 | 14.635 | 9.219 | 3.594 |
| 21 | AcNH$_2$···AcNH$_2$ | 15.274 | 16.299 | 16.472 | 16.094 | 15.380 | 12.460 | 8.007 | 3.008 |
| 22 | AcOH···Uracil | 18.324 | 19.524 | 19.759 | 19.361 | 18.565 | 15.198 | 9.867 | 4.150 |
| 23 | AcNH$_2$···Uracil | 18.076 | 19.195 | 19.432 | 19.086 | 18.367 | 15.265 | 10.240 | 4.662 |
| **π stack** | | | | | | | | | |
| 24 | Benzene···Benzene (π-π) | -0.138 | 1.791 | 2.521 | 2.633 | 2.450 | 1.479 | 0.468 | 0.059 |
| 25 | Pyridine···Pyridine (π-π) | 0.964 | 2.891 | 3.610 | 3.689 | 3.452 | 2.299 | 0.945 | 0.233 |
| 26 | Uracil···Uracil (π-π) | 7.593 | 9.266 | 9.619 | 9.256 | 8.541 | 5.993 | 3.071 | 0.986 |
| 27 | Benzene···Pyridine (π-π) | 0.362 | 2.413 | 3.156 | 3.234 | 3.001 | 1.913 | 0.705 | 0.143 |
| 28 | Benzene···Uracil (π-π) | 3.178 | 4.924 | 5.465 | 5.351 | 4.916 | 3.207 | 1.334 | 0.246 |
| 29 | Pyridine···Uracil (π-π) | 3.374 | 5.878 | 6.576 | 6.391 | 5.820 | 3.793 | 1.752 | 0.529 |
| 30 | Benzene···Ethene | 0.021 | 0.939 | 1.269 | 1.292 | 1.172 | 0.651 | 0.159 | -0.013 |
| 31 | Uracil···Ethene | 2.396 | 3.092 | 3.257 | 3.137 | 2.878 | 1.949 | 0.918 | 0.251 |
| 32 | Uracil···Ethyne | 2.587 | 3.387 | 3.587 | 3.464 | 3.183 | 2.159 | 1.012 | 0.264 |
| 33 | Pyridine···Ethene | 0.668 | 1.447 | 1.721 | 1.719 | 1.581 | 0.985 | 0.350 | 0.042 |
| **London dispersion complexes** | | | | | | | | | |
| 34 | Pentane···Pentane | 2.796 | 3.564 | 3.723 | 3.567 | 3.265 | 2.211 | 1.039 | 0.263 |
| 35 | Neopentane···Pentane | 1.822 | 2.460 | 2.580 | 2.456 | 2.228 | 1.479 | 0.690 | 0.185 |
| 36 | Neopentane···Neopentane | 1.425 | 1.705 | 1.742 | 1.625 | 1.502 | 1.018 | 0.497 | 0.132 |
| 37 | Cyclopentane···Neopentane | 1.576 | 2.217 | 2.373 | 2.295 | 2.111 | 1.449 | 0.699 | 0.189 |
| 38 | Cyclopentane···Cyclopentane | 2.198 | 2.800 | 2.956 | 2.824 | 2.563 | 1.693 | 0.780 | 0.201 |
| 39 | Benzene···Cyclopentane | 1.969 | 3.081 | 3.438 | 3.364 | 3.083 | 2.026 | 0.884 | 0.189 |
| 40 | Benzene···Neopentane | 1.712 | 2.549 | 2.791 | 2.721 | 2.502 | 1.674 | 0.762 | 0.187 |
| 41 | Uracil···Pentane | 3.679 | 4.560 | 4.724 | 4.502 | 3.999 | 2.402 | 0.961 | 0.212 |
| 42 | Uracil···Cyclopentane | 2.900 | 3.836 | 4.019 | 3.823 | 3.459 | 2.249 | 1.000 | 0.246 |
| 43 | Uracil···Neopentane | 2.756 | 3.500 | 3.613 | 3.415 | 3.083 | 2.013 | 0.912 | 0.229 |
| 44 | Ethene···Pentane | 1.597 | 1.918 | 1.943 | 1.822 | 1.639 | 1.069 | 0.481 | 0.117 |
| 45 | Ethyne···Pentane | 0.992 | 1.510 | 1.646 | 1.589 | 1.444 | 0.932 | 0.400 | 0.092 |
| 46 | Peptide···Pentane | 3.655 | 4.115 | 4.162 | 3.974 | 3.665 | 2.586 | 1.172 | 0.282 |
| **Mixed influence complexes** | | | | | | | | | |
| 47 | Benzene···Benzene (TS) | 1.493 | 2.452 | 2.762 | 2.731 | 2.536 | 1.734 | 0.818 | 0.222 |
| 48 | Pyridine···Pyridine (TS) | 2.402 | 3.199 | 3.440 | 3.375 | 3.154 | 2.259 | 1.168 | 0.370 |
| 49 | Benzene···Pyridine (TS) | 1.945 | 2.922 | 3.226 | 3.176 | 2.955 | 2.068 | 1.042 | 0.332 |
| 50 | Benzene···Ethyne (CH-π) | 1.791 | 2.580 | 2.810 | 2.747 | 2.545 | 1.772 | 0.890 | 0.272 |
| 51 | Ethyne···Ethyne (TS) | 1.170 | 1.434 | 1.489 | 1.431 | 1.318 | 0.913 | 0.452 | 0.134 |
| 52 | Benzene···AcOH (OH-π) | 3.908 | 4.525 | 4.655 | 4.508 | 4.213 | 3.107 | 1.697 | 0.554 |
| 53 | Benzene···AcNH$_2$ (NH-π) | 3.768 | 4.242 | 4.334 | 4.201 | 3.945 | 2.960 | 1.641 | 0.481 |
| 54 | Benzene···Water (OH-π) | 2.745 | 3.161 | 3.224 | 3.094 | 2.870 | 2.089 | 1.154 | 0.417 |
| 55 | Benzene···MeOH (OH-π) | 3.338 | 3.950 | 4.106 | 3.998 | 3.749 | 2.774 | 1.524 | 0.518 |
| 56 | Benzene···MeNH$_2$ (NH-π) | 2.381 | 2.977 | 3.139 | 3.044 | 2.803 | 1.922 | 0.933 | 0.260 |
| 57 | Benzene···Peptide (NH-π) | 3.595 | 4.828 | 5.176 | 5.059 | 4.715 | 3.384 | 1.799 | 0.619 |
| 58 | Pyridine···Pyridine (CH-N) | 2.884 | 3.869 | 4.166 | 3.890 | 3.442 | 2.173 | 1.005 | 0.274 |
| 59 | Ethyne···Water (CH-O) | 2.604 | 2.865 | 2.899 | 2.799 | 2.628 | 1.999 | 1.178 | 0.461 |
| 60 | Ethyne···AcOH (OH-π) | 4.306 | 4.787 | 4.854 | 4.679 | 4.371 | 3.234 | 1.760 | 0.556 |

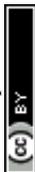

**Table 1** (continued)

| | Dimers | $0.90r_e$ | $0.95r_e$ | $1.00r_e$ | $1.05r_e$ | $1.10r_e$ | $1.25r_e$ | $1.50r_e$ | $2.00r_e$ |
|---|---|---|---|---|---|---|---|---|---|
| 61 | Pentane···AcOH | 2.629 | 2.835 | 2.820 | 2.677 | 2.467 | 1.757 | 0.773 | 0.166 |
| 62 | Pentane···AcNH₂ | 3.074 | 3.426 | 3.437 | 3.259 | 2.988 | 2.078 | 1.025 | 0.268 |
| 63 | Benzene···AcOH | 2.547 | 3.444 | 3.671 | 3.546 | 3.253 | 2.195 | 1.007 | 0.259 |
| 64 | Peptide···Ethene | 2.527 | 2.881 | 2.928 | 2.801 | 2.585 | 1.828 | 0.867 | 0.188 |
| 65 | Pyridine···Ethyne | 3.661 | 4.008 | 4.066 | 3.946 | 3.724 | 2.866 | 1.679 | 0.618 |
| 66 | MeNH₂···Pyridine | 3.360 | 3.796 | 3.893 | 3.784 | 3.560 | 2.677 | 1.488 | 0.490 |

[a] Sterling silver = MP2-F12/haV{T,Q}Z-F12 full-CP + [CCSD(F12*) − MP2-F12]/VTZ-F12 "raw" + (T)/sano-PV{D,T}Z+ "raw".

**Table 2** Root-mean-square deviations (RMSDs, kcal mol⁻¹) of Hobza's original, Martin's revised (ref. 42), and "bronze" standard S66x8 dissociation energies evaluated relative to the "sterling silver" reference. Together with full S66x8, we have also included the RMSDs for its four subsets (*i.e.*, hydrogen bonds, π-stacking, London dispersion complexes, and mixed-influence interactions). Heat mapping is from red (worst) *via* yellow to green (best)

| Old reference dissociation energies of S66x8 | | RMSD (kcal/mol) | | | | | | | | |
|---|---|---|---|---|---|---|---|---|---|---|
| | | **Full S66x8** | **Intermolecular Distances** | | | | | | | |
| | | | **$0.90r_e$** | **$0.95r_e$** | **$r_e$** | **$1.05r_e$** | **$1.10r_e$** | **$1.25r_e$** | **$1.50r_e$** | **$2.00r_e$** |
| Hobza[37] | Full S66x8 | 0.103 | 0.183 | 0.148 | 0.118 | 0.094 | 0.074 | 0.035 | 0.012 | 0.003 |
| | Hydrogen Bonds | 0.111 | 0.203 | 0.161 | 0.125 | 0.096 | 0.072 | 0.027 | 0.006 | 0.004 |
| | π-stack | 0.168 | 0.299 | 0.240 | 0.192 | 0.154 | 0.123 | 0.062 | 0.021 | 0.005 |
| | London Dispersion | 0.069 | 0.109 | 0.096 | 0.082 | 0.071 | 0.058 | 0.033 | 0.012 | 0.003 |
| | Mixed Influence | 0.062 | 0.104 | 0.088 | 0.073 | 0.061 | 0.050 | 0.027 | 0.009 | 0.002 |
| Brauer[42] | Full S66x8 | 0.096 | 0.131 | 0.125 | 0.117 | 0.108 | 0.098 | 0.071 | 0.039 | 0.014 |
| | Hydrogen Bonds | 0.059 | 0.060 | 0.064 | 0.068 | 0.070 | 0.071 | 0.063 | 0.041 | 0.015 |
| | π-stack | 0.171 | 0.236 | 0.225 | 0.209 | 0.191 | 0.172 | 0.117 | 0.061 | 0.022 |
| | London Dispersion | 0.102 | 0.150 | 0.138 | 0.124 | 0.111 | 0.096 | 0.064 | 0.032 | 0.012 |
| | Mixed Influence | 0.072 | 0.101 | 0.095 | 0.087 | 0.080 | 0.071 | 0.049 | 0.027 | 0.010 |
| Bronze[43] | Full S66x8 | 0.041 | 0.054 | 0.050 | 0.049 | 0.047 | 0.044 | 0.035 | 0.020 | 0.007 |
| | Hydrogen Bonds | 0.033 | 0.062 | 0.044 | 0.031 | 0.023 | 0.020 | 0.023 | 0.018 | 0.006 |
| | π-stack | 0.079 | 0.089 | 0.093 | 0.098 | 0.097 | 0.092 | 0.067 | 0.036 | 0.012 |
| | London Dispersion | 0.033 | 0.032 | 0.039 | 0.042 | 0.043 | 0.039 | 0.030 | 0.016 | 0.006 |
| | Mixed Influence | 0.024 | 0.025 | 0.027 | 0.029 | 0.029 | 0.029 | 0.022 | 0.013 | 0.005 |

upper block of Table 3). Hence, the more economical of the two would seem preferable.

Let us take a closer look at the composite LNO-CCSD(T) schemes (a.k.a. cLNOs[62]) with different accuracy thresholds and basis sets. Regardless of CP correction or lack thereof, the RMS deviations of all the cLNO methods are below 0.1 kcal mol⁻¹ for the complete S66x8 set. With RMS error of 0.024 and 0.028 kcal mol⁻¹, respectively, vTight{T,Q} + 0.31[vvTight − vTight]/T half-CP and vTight{Q,5} + 0.71[vvTight − vTight]/T half-CP are the two best picks among the eleven cLNOs listed in Table 3. However, the first cLNO scheme is clearly preferred due to the lower computational cost. Either with "raw" or half-CP, the most affordable composite scheme Normal{T,Q} + $c_1$[vTight − Normal]/T performs similarly to the very expensive LNO-CCSD(T, vTight)/haV5Z half-CP.

In special cases, where a counterpoise correction is impossible (*e.g.*, intramolecular or conformer interactions), the composite method, vTight{Q,5} + 1.55[vvTight − vTight]/T offers excellent performance (0.030 kcal mol⁻¹). Another low-cost alternative for such situations is the Tight{T,Q} + 0.72[vvTight − Tight]/T scheme (0.039 kcal mol⁻¹). Without CP corrections, the two-tier scheme, Tight{T,Q} + 0.72[vvTight − Tight]/T, marginally outperforms more expensive three-tier method,

Tight{Q,5} + 2.45[vvTight − vTight]/T + 0.59[vTight − Tight]/Q, which is contrary to what we observed for S66 in a previous study.[62] With half-CP, a relatively low-cost method, Tight{T,Q} + 0.95[vTight − Tight]/T, offers very good accuracy (0.037 kcal mol⁻¹) when compared to the "sterling silver" level S66x8 interaction energies.

#### (c) PNO-LCCSD(T)-based methods

The RMS deviations for the pure and composite PNO-LCCSD(T) methods using different basis sets and accuracy thresholds are listed in Table 4.

Overall, the counterpoise uncorrected results indicate a consistent improvement in accuracy with increasing basis set size for any threshold. Among the PNO-LCCSD(T)/haV$n$Z methods, PNO-LCCSD(T, Tight)/haVQZ offers the best performance (0.071 kcal mol⁻¹), followed by PNO-LCCSD(T, Default)/haV5Z (0.075 kcal mol⁻¹). Further investigation reveals that the former method does better for the hydrogen-bonded systems and London dispersion complexes, whereas the latter is preferred for the π-stacks. Using a full-CP correction worsens performance across the board. However, half-CP can still perform close to the "raw" accuracy with a large basis set. Now, comparing the performance of "raw" and half-CP variants of

**Table 3** Root-mean-square deviations (RMSDs, kcal mol⁻¹) of pure and composite LNO-CCSD(T) methods with respect to the "sterling silver" reference. Heat mapping is from red (worst) via yellow to green (best)

| Threshold | Basis set | RMSD (kcal/mol) | | | | | | | | | | | | | | |
| | | Total | | | Hydrogen bonds | | | π stack | | | London Dispersion | | | Mixed Influence | | |
| | | Raw | CP | Half | Raw | CP | Half | Raw | CP | Half | Raw | CP | Half | Raw | CP | Half |
| Normal | haVTZ | 0.624 | 0.357 | 0.391 | 0.123 | 0.558 | 0.300 | 1.279 | 0.253 | 0.731 | 0.584 | 0.128 | 0.331 | 0.479 | 0.142 | 0.248 |
| | haVQZ | 0.337 | 0.237 | 0.268 | 0.108 | 0.293 | 0.194 | 0.700 | 0.347 | 0.518 | 0.297 | 0.140 | 0.214 | 0.244 | 0.118 | 0.174 |
| | haVSZ | 0.261 | 0.209 | 0.232 | 0.150 | 0.199 | 0.172 | 0.528 | 0.371 | 0.449 | 0.206 | 0.150 | 0.177 | 0.178 | 0.123 | 0.149 |
| | haV{T,Q}Z | 0.169 | 0.221 | 0.193 | 0.133 | 0.131 | 0.131 | 0.328 | 0.436 | 0.381 | 0.127 | 0.172 | 0.144 | 0.101 | 0.166 | 0.133 |
| | haV{Q,5}Z | 0.206 | 0.202 | 0.198 | 0.206 | 0.112 | 0.155 | 0.356 | 0.403 | 0.378 | 0.132 | 0.169 | 0.146 | 0.127 | 0.142 | 0.130 |
| Tight | haVTZ | 0.515 | 0.320 | 0.268 | 0.111 | 0.492 | 0.228 | 1.030 | 0.195 | 0.477 | 0.501 | 0.133 | 0.236 | 0.408 | 0.171 | 0.165 |
| | haVQZ | 0.208 | 0.150 | 0.140 | 0.047 | 0.231 | 0.130 | 0.425 | 0.111 | 0.245 | 0.192 | 0.066 | 0.110 | 0.160 | 0.061 | 0.086 |
| | haVSZ | 0.123 | 0.106 | 0.104 | 0.071 | 0.149 | 0.109 | 0.245 | 0.115 | 0.174 | 0.096 | 0.063 | 0.072 | 0.087 | 0.046 | 0.061 |
| | haV{T,Q}Z | 0.061 | 0.087 | 0.065 | 0.078 | 0.067 | 0.072 | 0.069 | 0.151 | 0.101 | 0.050 | 0.068 | 0.047 | 0.035 | 0.071 | 0.042 |
| | haV{Q,5}Z | 0.085 | 0.084 | 0.079 | 0.118 | 0.070 | 0.091 | 0.085 | 0.141 | 0.109 | 0.068 | 0.076 | 0.067 | 0.037 | 0.063 | 0.045 |
| vTight | haVTZ | 0.492 | 0.297 | 0.224 | 0.144 | 0.441 | 0.177 | 0.977 | 0.191 | 0.419 | 0.468 | 0.160 | 0.186 | 0.392 | 0.181 | 0.137 |
| | haVQZ | 0.177 | 0.112 | 0.092 | 0.048 | 0.169 | 0.071 | 0.363 | 0.057 | 0.179 | 0.158 | 0.069 | 0.070 | 0.138 | 0.055 | 0.056 |
| | haVSZ | 0.081 | 0.063 | 0.056 | 0.017 | 0.092 | 0.049 | 0.171 | 0.047 | 0.105 | 0.065 | 0.048 | 0.038 | 0.064 | 0.027 | 0.035 |
| | haV{T,Q}Z | 0.042 | 0.045 | 0.025 | 0.028 | 0.021 | 0.019 | 0.051 | 0.086 | 0.029 | 0.059 | 0.038 | 0.037 | 0.037 | 0.039 | 0.018 |
| | haV{Q,5}Z | 0.044 | 0.044 | 0.033 | 0.047 | 0.025 | 0.033 | 0.051 | 0.082 | 0.039 | 0.053 | 0.039 | 0.042 | 0.028 | 0.038 | 0.023 |
| vvTight | haVTZ | 0.496 | 0.281 | 0.220 | 0.168 | 0.418 | 0.158 | 0.979 | 0.178 | 0.421 | 0.473 | 0.148 | 0.187 | 0.394 | 0.173 | 0.138 |

| | Coefficients | | | | Composite Methods (cLNO) | | | | | | | | | | | |
| | | $c_1$ | | | $c_2$ | | | | | | | | | | | |
| | | Raw | CP | Half | | | | | | | | | | | | |
| Normal{T,Q} + $c_1$[Tight − Normal]/T | | 0.93 | 1.06 | 1.02 | ------ | 0.057 | 0.083 | 0.059 | 0.037 | 0.066 | 0.044 | 0.068 | 0.115 | 0.077 | 0.085 | 0.093 | 0.078 | 0.044 | 0.075 | 0.048 |
| Normal{Q,5} + $c_1$[Tight − Normal]/T | | 1.09 | 0.96 | 1.03 | ------ | 0.085 | 0.082 | 0.068 | 0.084 | 0.068 | 0.055 | 0.080 | 0.113 | 0.078 | 0.095 | 0.094 | 0.083 | 0.083 | 0.068 | 0.066 |
| Tight{T,Q} + $c_1$[vTight − Tight]/T | | 0.79 | 1.04 | 0.95 | ------ | 0.045 | 0.057 | 0.037 | 0.040 | 0.034 | 0.030 | 0.048 | 0.093 | 0.046 | 0.061 | 0.058 | 0.050 | 0.037 | 0.052 | 0.030 |
| Tight{T,Q} + $c_1$[vvTight − Tight]/T | | 0.72 | 0.78 | 0.78 | ------ | 0.039 | 0.061 | 0.035 | 0.029 | 0.042 | 0.028 | 0.044 | 0.106 | 0.052 | 0.045 | 0.047 | 0.037 | 0.034 | 0.056 | 0.036 |
| Tight{Q,5} + $c_1$[vTight − Tight]/T | | 1.26 | 0.84 | 1.05 | ------ | 0.056 | 0.065 | 0.053 | 0.054 | 0.038 | 0.041 | 0.052 | 0.094 | 0.053 | 0.079 | 0.093 | 0.084 | 0.038 | 0.048 | 0.036 |
| Tight{Q,5} + $c_1$[vvTight − Tight]/Q | | 1.14 | 0.89 | 1.07 | ------ | 0.060 | 0.064 | 0.051 | 0.061 | 0.054 | 0.037 | 0.052 | 0.090 | 0.050 | 0.083 | 0.082 | 0.040 | 0.049 | 0.037 | |
| Tight{Q,5} + $c_1$[vvTight − Tight]/T | | 1.11 | 0.67 | 0.9 | ------ | 0.045 | 0.065 | 0.047 | 0.038 | 0.042 | 0.034 | 0.043 | 0.102 | 0.055 | 0.068 | 0.080 | 0.070 | 0.034 | 0.051 | 0.035 |
| Tight{Q,5} + $c_1$[vTight − vTight]/T | | 3.28 | 0.68 | 1.81 | ------ | 0.051 | 0.080 | 0.064 | 0.049 | 0.060 | 0.060 | 0.071 | 0.139 | 0.101 | 0.054 | 0.070 | 0.054 | 0.038 | 0.065 | 0.047 |
| Tight{Q,5} + $c_1$[vvTight − vTight]/T + $c_2$[vTight − Tight]/Q | | 2.45 | 0.22 | 0.95 | 0.59 (Raw) 0.86 (CP) 0.88 (Half) | 0.043 | 0.063 | 0.046 | 0.038 | 0.035 | 0.034 | 0.047 | 0.092 | 0.054 | 0.059 | 0.089 | 0.068 | 0.034 | 0.049 | 0.036 |
| vTight{T,Q} + $c_1$[vvTight − vTight]/T | | 0.77 | -0.02 | 0.31 | ------ | 0.039 | 0.045 | 0.024 | 0.019 | 0.021 | 0.019 | 0.052 | 0.086 | 0.028 | 0.053 | 0.038 | 0.035 | 0.036 | 0.039 | 0.018 |
| vTight{Q,5} + $c_1$[vvTight − vTight]/T | | 1.55 | 0.12 | 0.71 | ------ | 0.030 | 0.044 | 0.028 | 0.019 | 0.024 | 0.023 | 0.039 | 0.083 | 0.035 | 0.043 | 0.038 | 0.035 | 0.025 | 0.038 | 0.023 |

**Table 4** Root-mean-square deviations (RMSDs, kcal mol⁻¹) of the pure and composite PNO-LCCSD(T) methods with respect to the "sterling silver" reference. Heat mapping is from red (worst) via yellow to green (best)

| Threshold | Basis Set | RMSD (kcal/mol) | | | | | | | | | | | | | | |
| | | Total | | | Hydrogen bonds | | | π stack | | | London Dispersion | | | Mixed Influence | | |
| | | Raw | CP | Half | Raw | CP | Half | Raw | CP | Half | Raw | CP | Half | Raw | CP | Half |
| Default | haVTZ | 0.132 | 0.428 | 0.258 | 0.180 | 0.550 | 0.363 | 0.132 | 0.438 | 0.197 | 0.105 | 0.333 | 0.204 | 0.069 | 0.299 | 0.150 |
| | haVQZ | 0.098 | 0.176 | 0.136 | 0.120 | 0.225 | 0.172 | 0.112 | 0.183 | 0.147 | 0.094 | 0.139 | 0.116 | 0.059 | 0.115 | 0.086 |
| | haVSZ | 0.075 | 0.099 | 0.086 | 0.089 | 0.129 | 0.109 | 0.095 | 0.106 | 0.101 | 0.064 | 0.074 | 0.069 | 0.046 | 0.061 | 0.053 |
| | haV{T,Q}Z | 0.131 | 0.027 | 0.072 | 0.085 | 0.021 | 0.051 | 0.236 | 0.042 | 0.130 | 0.118 | 0.028 | 0.068 | 0.053 | 0.024 | 0.051 |
| | haV{Q,5}Z | 0.056 | 0.028 | 0.040 | 0.053 | 0.035 | 0.047 | 0.083 | 0.037 | 0.058 | 0.040 | 0.022 | 0.028 | 0.037 | 0.018 | 0.025 |
| Tight | haVTZ | 0.299 | 0.362 | 0.163 | 0.061 | 0.486 | 0.237 | 0.651 | 0.315 | 0.172 | 0.220 | 0.278 | 0.091 | 0.219 | 0.247 | 0.055 |
| | haVQZ | 0.071 | 0.143 | 0.072 | 0.037 | 0.190 | 0.109 | 0.156 | 0.137 | 0.028 | 0.047 | 0.111 | 0.052 | 0.041 | 0.094 | 0.033 |
| | haV{T,Q}Z | 0.103 | 0.020 | 0.055 | 0.062 | 0.015 | 0.033 | 0.189 | 0.032 | 0.106 | 0.092 | 0.021 | 0.056 | 0.084 | 0.019 | 0.040 |

| Coefficients ($c_1$) | | | | Composite Methods (cPNO) | | | | | | | | | | | | |
| | Raw | CP | Half | | | | | | | | | | | | | |
| Default{T,Q} + $c_1$[Tight − Default]/T | 0.39 | 0.07 | 0.34 | 0.054 | 0.027 | 0.035 | 0.031 | 0.019 | 0.021 | 0.101 | 0.041 | 0.066 | 0.045 | 0.027 | 0.031 | 0.047 | 0.026 | 0.025 |
| Default{Q,5} + $c_1$[Tight − Default]/T | 0.15 | 0.2 | 0.16 | 0.044 | 0.024 | 0.027 | 0.039 | 0.025 | 0.030 | 0.043 | 0.033 | 0.039 | 0.026 | 0.020 | 0.022 | 0.022 | 0.020 | 0.019 |
| Default{Q,5} + $c_1$[Tight − Default]/Q | 0.33 | 0.47 | 0.36 | 0.033 | 0.024 | 0.027 | 0.038 | 0.024 | 0.029 | 0.047 | 0.031 | 0.037 | 0.026 | 0.021 | 0.023 | 0.021 | 0.021 | 0.019 |
| Tight{T,Q} + $c_1$[Tight − Default]/T | 0.3 | 0.02 | 0.25 | 0.049 | 0.020 | 0.029 | 0.029 | 0.009 | 0.016 | 0.090 | 0.030 | 0.053 | 0.043 | 0.021 | 0.026 | 0.042 | 0.019 | 0.023 |
| Tight{T,Q} + $c_1$[Tight − Default]/Q | 0.66 | 0.03 | 0.56 | 0.049 | 0.020 | 0.034 | 0.030 | 0.015 | 0.021 | 0.088 | 0.031 | 0.051 | 0.044 | 0.020 | 0.026 | 0.041 | 0.019 | 0.021 |

PNO-LCCSD(T, Tight)/haVQZ for the four subsets of S66x8, we have found that for the hydrogen bonding complexes, "raw" is preferred, but for the π-stack dimers, "half-CP" variant offers significantly better accuracy than its half-CP counterpart. For the London dispersion and mixed influence dimers, there is very little to choose between these two methods (see Table 4).

Now, what is the effect of CBS extrapolation? Both haV{T,Q}Z and haV{Q,5}Z, in combination with the full-CP and Default threshold, perform remarkably well (0.027 and 0.028 kcal mol⁻¹, respectively). Using a tighter threshold marginally improves the accuracy of PNO-LCCSD(T)/haV{T,Q}Z (0.020 kcal mol⁻¹). Hence, irrespective of the choice of accuracy threshold, the standard PNO-LCCSD(T)/CBS methods with full-CP perform similar to the best cLNO schemes.

Unlike for cLNOs, we have not found any significant improvement in accuracy when PNO-LCCSD(T)-based composite schemes (i.e., cPNOs) are considered instead of the standard PNO-LCCSD(T)/haV$nZ$ or PNO-LCCSD(T)/CBS methods with full-CP. However, for special cases where counterpoise correction is not applicable (in other words using "raw" energies), the composite methods outperform the standard PNO-LCCSD(T) alternatives. With RMS deviations of 0.033 and 0.034 kcal mol⁻¹, respectively, Default{Q,5} + 0.15[Tight − Default]/T and Default{Q,5} + 0.33[Tight − Default]/Q are the two best picks in the counterpoise uncorrected category. In the same vein, comparing the top performers from the "raw" cLNO and cPNO schemes, we have found that for hydrogen bonding and $\pi$-stack complexes, cLNOs are preferred, whereas cPNOs offer better accuracy for London dispersion complexes. For the mixed influence subset, there is very little to choose between cLNO and cPNO.

### (d) DLPNO-CCSD($T_1$)-based methods

Table 5 summarizes the RMSDs of the pure and composite DLPNO-CCSD($T_1$) methods for different basis sets, accuracy thresholds, and $T_{cutPNO}$ combinations.

The "raw" results indicate a gradual improvement of accuracy with increasing basis set size and tightening of the thresholds, provided that we use the default $T_{cutPNO}$ (i.e., $3.33 \times 10^{-7}$ and $1.0 \times 10^{-7}$ for Normal and TightPNO, respectively). Moving from "raw" to full-CP correction further improves the performance of DLPNO-CCSD($T_1$, Normal)/haVTZ, but stays more or less indifferent when haVQZ is employed. However, while using the Tight threshold, the performance of "raw" and full-CP DLPNO-CCSD($T_1$)/haVTZ are comparable, but with haVQZ basis set, "raw" performs better.

Fortuitously, the strategy of using the threshold $T_{CutPNO} = 1.0 \times 10^{-6}$ in TightPNO and a haVTZ basis set offers the best performance (0.309 kcal mol$^{-1}$) for "raw" interaction energies. The RMS error counter-intuitively increases as the $T_{cutPNO}$ parameter becomes tighter. When CP correction is included, we observe the reverse trend. Irrespective of the choice of $T_{cutPNO}$, half-CP correction is more beneficial than full-CP. With the same basis set, tightening the accuracy threshold further (i.e., VeryTightPNO) is only beneficial for half-CP. Using a more extensive basis set and default $T_{cutPNO}$ improves accuracy throughout, with and without CP correction. Although "raw" DLPNO-CCSD($T_1$, Tight)/haVTZ underperforms "raw" DLPNO-CCSD($T_1$, Tight)/haVQZ for $\pi$-stacks, London dispersion, and mixed influence subsets, these two methods offer similar performance for H-bonds.

Now, what is the effect of CBS extrapolation? Except for the hydrogen bonding, we noticed an improvement in performance across the board when the Normal threshold is employed. By tightening the accuracy threshold further, we found that CBS extrapolation does more harm than good for "raw" but improves the performance for full and half-CP. With full-CP correction, the DLPNO-CCSD($T_1$, Tight)/haV{T,Q}Z (0.095 kcal mol$^{-1}$) is the best pick among the pure methods, followed by its half-CP version (0.117 kcal mol$^{-1}$). Using DLPNO-CCSD($T_1$, Tight)/haVQZ/$T_{cutPNO} = 10^{-7}$ and DLPNO-CCSD($T_1$, Tight)/haVTZ/CPS interaction energies for two-point CBS extrapolation worsen the performance regardless of considering a CP correction.

Among the counterpoise-uncorrected DLPNO-CCSD($T_1$)-based composite schemes, a relatively low-cost composite method, Normal/{T,Q} + 0.93[Tight/CPS − Normal]/T offers the best accuracy (0.072 kcal mol$^{-1}$). The use of counterpoise correction only offers marginal improvement (RMSD = 0.065 and 0.067 kcal mol$^{-1}$, with full and half-CP, respectively). With an RMS error of 0.059 kcal mol$^{-1}$, the full CP corrected composite method, Tight{T,Q} + 1.1[vTight − Tight]/T is the best pick among all the standard and composite DLPNO-CCSD($T_1$) listed in Table 5.

DLPNO-CCSD($T_1$) is much more demanding in terms of I/O storage and bandwidth requirements than DLPNO-CCSD($T_0$), and this will hold especially true for the largest basis sets. It was previously observed by Iron and Janes[57,58] and by Efremenko and Martin,[55] both in the context of organometallic catalysis, that the ($T_1$)–($T_0$) difference is relatively insensitive to the basis set; hence, we considered here a two-tier method, DLPNO-CCSD($T_0$)/haVQZ + $c_1$[DLPNO-CCSD($T_0$)/haVQZ − DLPNO-CCSD($T_0$)/haVTZ] + $c_2$[DLPNO-CCSD($T_1$)/haVTZ − DLPNO-CCSD($T_0$)/haVTZ], where the CBS extrapolation is carried out at the DLPNO-CCSD($T_0$) level and the ($T_1$) − ($T_0$) difference is evaluated in a smaller basis set. With a root-mean-square error of 0.079 kcal mol$^{-1}$, the performance of this cDLPNO-scheme is comparable to the best pick in the "raw" category, Normal/{T,Q} + 0.93[Tight/CPS − Normal]/T. While using counterpoise-uncorrected energies, the optimized coefficient for the ($T_1$) − ($T_0$) contribution is anomalously large ($c_2 = 3.33$). However, with full-CP correction, $c_2$ is reduced to 1.33, and the RMS error improves to 0.068 kcal mol$^{-1}$, which is not very far from the accuracy of the more expensive DLPNO-CCSD($T_1$)-based composite method, Tight{T,Q} + 1.1[vTight − Tight]/T (see Table 5). These two-tier cDLPNO methods may be an attractive option for larger systems.

### (e) A few observations regarding computational requirements

This is not the appropriate place to enter into a discussion on the relative computational efficiency of the various codes—particularly since our computing cluster is extremely heterogeneous in terms of CPUs, and a fair comparison would require not only "all else being equal" but a broad and representative sample.

That being said, we can state a few observations concerning the effect of different accuracy settings. For illustration, wall clock timings for system 26 (stacked uracil dimer) have been given in the ESI.† All these calculations were carried out on 8 cores of an otherwise idle 2.4GHz Intel Xeon Gold Intel(R) Xeon(R) Gold 6240R "Cascade Lake" node with 192GB of RAM—the ORCA jobs, which are more demanding in terms of I/O bandwidth, were run on a machine with the same CPU and 2.9TB of striped SSD.

As a rule, for LNO-CCSD(T), turning up accuracy from Normal to Tight requires between triple and quintuple the wall clock time.

The additional expense going from Tight to vTight tends to be smaller—doubling to tripling wall clock time.

As for the dependence of wall clock times on the basis set size, naively one would expect it to tend towards linear scaling with the number of basis functions for sufficiently large systems and basis sets. In practice, we are not yet operating in such a size regime: the numbers of basis functions are in a 848 : 408 = 2.08 ratio for haVTZ : haVDZ, and a 1520 : 848 = 1.79 ratio for haVQZ : haVTZ, but the wall clock times definitely go up faster than that, by a factor of 3–4. Still, this is much gentler than the $N^4$ scaling of the canonical calculation—so while canonical CCSD(T) calculations in the smallest haVDZ basis set may not be abysmally more demanding than their linearized counterpart, a wide chasm opens for the larger basis set. This militates in favor of a composite method where basis set extrapolation is carried out using localized methods, and the difference between canonical and localized method in the smallest basis set used as a differential correction.

**Table 5** Root-mean-square deviations (RMSDs, kcal mol$^{-1}$) of the pure and composite DLPNO-CCSD(T$_1$) methods with respect to the "sterling silver" reference. Heat mapping is from red (worst) *via* yellow to green (best)

| Threshold | Basis set | T$_{cutPNO}$ | Total Raw | Total CP | Total Half | H-bonds Raw | H-bonds CP | H-bonds Half | π stack Raw | π stack CP | π stack Half | London Dispersion Raw | London Dispersion CP | London Dispersion Half | Mixed Influence Raw | Mixed Influence CP | Mixed Influence Half |
|---|---|---|---|---|---|---|---|---|---|---|---|---|---|---|---|---|---|
| Normal | haVTZ | default (i.e., $3.33\times10^{-7}$) | 0.604 | 0.515 | 0.453 | 0.264 | 0.799 | 0.525 | 1.296 | 0.318 | 0.770 | 0.413 | 0.263 | 0.196 | 0.411 | 0.218 | 0.199 |
|  | haVQZ |  | 0.368 | 0.353 | 0.342 | 0.309 | 0.594 | 0.405 | 0.737 | 0.417 | 0.570 | 0.186 | 0.152 | 0.143 | 0.205 | 0.153 | 0.148 |
|  | haV{T,Q}Z |  | 0.267 | 0.286 | 0.275 | 0.345 | 0.316 | 0.330 | 0.392 | 0.500 | 0.446 | 0.133 | 0.126 | 0.124 | 0.106 | 0.140 | 0.121 |
| Tight | haVTZ | default (i.e., $1.0\times10^{-7}$) | 0.410 | 0.419 | 0.217 | 0.080 | 0.572 | 0.290 | 0.806 | 0.395 | 0.263 | 0.404 | 0.297 | 0.134 | 0.341 | 0.262 | 0.115 |
|  | haVTZ | $1.0\times10^{-6}$ | 0.309 | 0.563 | 0.283 | 0.154 | 0.716 | 0.428 | 0.600 | 0.616 | 0.191 | 0.268 | 0.445 | 0.163 | 0.246 | 0.371 | 0.139 |
|  | haVTZ | CPS on 1.0E−{6,7} | 0.471 | 0.351 | 0.220 | 0.111 | 0.500 | 0.223 | 0.913 | 0.289 | 0.355 | 0.476 | 0.227 | 0.177 | 0.392 | 0.210 | 0.141 |
|  | haVQZ | $1.0\times10^{-7}$ | 0.124 | 0.215 | 0.128 | 0.085 | 0.284 | 0.181 | 0.213 | 0.235 | 0.115 | 0.114 | 0.151 | 0.080 | 0.106 | 0.153 | 0.075 |
|  | haV{T,Q}Z | $1.0\times10^{-7}$ | 0.146 | 0.095 | 0.117 | 0.133 | 0.100 | 0.116 | 0.236 | 0.141 | 0.134 | 0.127 | 0.068 | 0.093 | 0.106 | 0.075 | 0.085 |
| VeryTight | haV{T,Q}Z | CPS on 1.0E−{6,7} | 0.191 | 0.136 | 0.162 | 0.176 | 0.146 | 0.161 | 0.306 | 0.203 | 0.252 | 0.174 | 0.106 | 0.138 | 0.136 | 0.093 | 0.112 |
|  | haVTZ | $1.0\times10^{-8}$ | 0.402 | 0.379 | 0.176 | 0.102 | 0.510 | 0.222 | 0.806 | 0.345 | 0.243 | 0.378 | 0.291 | 0.093 | 0.321 | 0.247 | 0.073 |

| Method | Coefficients ($c_i$) Raw | CP | Half | Total Raw | CP | Half | H-bonds Raw | CP | Half | π stack Raw | CP | Half | London Disp. Raw | CP | Half | Mixed Infl. Raw | CP | Half |
|---|---|---|---|---|---|---|---|---|---|---|---|---|---|---|---|---|---|---|
| Normal{T,Q} + $c_i$[Tight − Normal]/T | 0.90 | 0.94 | 0.92 | 0.104 | 0.091 | 0.096 | 0.124 | 0.101 | 0.112 | 0.148 | 0.134 | 0.139 | 0.076 | 0.073 | 0.073 | 0.058 | 0.057 | 0.056 |
| Normal{T,Q} + $c_i$[Tight/CPS − Normal]/T | 0.93 | 0.99 | 0.96 | 0.072 | 0.065 | 0.067 | 0.062 | 0.041 | 0.049 | 0.100 | 0.098 | 0.097 | 0.080 | 0.068 | 0.074 | 0.057 | 0.066 | 0.061 |
| $(T_0)$Tight/Q + $c_1$[($T_0$)Tight/Q − ($T_0$)Tight/T] + $c_2$[[T,Q]Tight/T − ($T_0$)Tight/T] | $c_1$=0.61 $c_2$=3.33 | $c_1$=0.93 $c_2$=13.3 | $c_1$=0.95 $c_2$=3.40 | 0.079 | 0.068 | 0.072 | 0.094 | 0.042 | 0.067 | 0.094 | 0.101 | 0.100 | 0.055 | 0.053 | 0.046 | 0.064 | 0.081 | 0.073 |
| Tight{T,Q} + $c_i$[Tight − Normal]/T | −0.11 | −0.03 | −0.08 | 0.143 | 0.097 | 0.116 | 0.162 | 0.110 | 0.137 | 0.185 | 0.131 | 0.149 | 0.128 | 0.067 | 0.090 | 0.074 | 0.074 | 0.081 |
| Tight{T,Q} + $c_i$[vTight − Tight]/T | 1.02 | 1.10 | 1.11 | 0.228 | 0.059 | 0.090 | 0.080 | 0.045 | 0.059 | 0.211 | 0.079 | 0.142 | 0.139 | 0.058 | 0.097 | 0.109 | 0.060 | 0.081 |
| Tight{T,Q}·CPS + $c_i$[Tight/CPS − Normal]/T | 0.05 | 0.04 | 0.03 | 0.191 | 0.136 | 0.162 | 0.161 | 0.134 | 0.151 | 0.325 | 0.219 | 0.265 | 0.174 | 0.106 | 0.137 | 0.137 | 0.094 | 0.112 |

Another observation that may be relevant here concerns the PNO cutoff extrapolation of the $T_{cutPNO} = 1.0 \times 10^{-\{6,7\}}$ type. The more expensive $T_{cutPNO} = 1.0 \times 10^{-7}$ calculation is equivalent to standard Tight, while the $T_{cutPNO} = 1.0 \times 10^{-6}$ calculation is not much more expensive than Normal. So, it would seem that this is an economical as well as accurate option.

### (f) How far can we go with canonical CCSD(T)?

We could do the canonical CCSD(T) calculations with density fitting and haV$nZ$ ($n = $ D and T) basis set for the whole S66x8 set. Relative to the "sterling silver" reference, DF-CCSD(T)/haVDZ offers RMS deviations of 1.16, 0.90, and 0.53 kcal mol$^{-1}$ with "raw", full-CP, and half-CP correction, respectively. Except for NCIs involving hydrogen bonds, CP correction is beneficial across the board. Using a larger basis set improves RMSDs substantially, and on top of that, a two-point CBS extrapolation (using Schwenke's[100] formula) further ameliorates the statistics. With full-CP, DF-CCSD(T)/haV{T,T}Z offers an RMS deviation of 0.11 kcal mol$^{-1}$. For $\pi$-stack, London dispersion, and mixed influence subsets, full-CP performs noticeably better than the "raw" and half-CP. The only exceptions are the H-bonded systems, where half-CP wins the race. Compared to the "sterling silver" level HLC, the canonical [CCSD(T)-MP2]/haVTZ energies offer RMS errors of 0.04, 0.05, and 0.07 kcal mol$^{-1}$ with "raw", full, and half CP correction.

Now, let us compare the performance of different localized coupled cluster methods relative to the canonical DF-CCSD(T) interaction energies. LNO-CCSD(T) with vTight and vvTight settings perform remarkably if we do not use any CP correction. Irrespective of the choice of accuracy threshold, full-CP correction is beneficial for the pure PNO-CCSD(T) methods, but they are not even close to "raw" LNO-CCSD(T)/vTight. However, with full CP correction, the PNO-based composite method, Tight + A(Tight − Default), offers accuracy similar to "raw" LNO-CCSD(T)/vTight. For DLPNO-CCSD(T$_1$), it does not make a significant difference whether we use counterpoise correction or not (see Table 6).

### (g) Some remarks on the evaluation of more approximate methods, such as DFT functionals and SAPT(DFT)

The revised reference data prompt the question: to what extent do the revised values affect or modify prior observations (*e.g.*, in ref. 42) on the performance of DFT functionals for the S66x8 dataset?

The easiest way to see this would be to take the Excel workbook in the ESI† of the said paper, splice in our new reference data, and compare the published tables in the paper with the dynamic versions in leaf "Summary" of the ESI,† particularly for Tables 14 and 15.

Generally speaking, the conclusions from ref. 42 are unaffected: for instance, the RMSDs for BLYP-D3BJ and BP86-D3BJ with the def2-QZVP[101] basis set and full counterpoise change from 0.23 and 0.58 kcal mol$^{-1}$, respectively, to 0.22 and 0.65 kcal mol$^{-1}$. This does not affect the superiority of BLYP over BP86 in this context. (The corresponding changes for B3LYP-D3BJ and PBE0-D3BJ are from 0.20 and 0.35 kcal mol$^{-1}$, respectively, to 0.23 and 0.36 kcal mol$^{-1}$.) Among double hybrids, all with haVQZ basis set and half-counterpoise, B2PLYP-D3BJ[102] and B2GP-PLYP-D3BJ[103] actually improve from 0.19 and 0.22 kcal mol$^{-1}$ to 0.15 and 0.18 kcal mol$^{-1}$, respectively, while DSD-PBEP86-D3BJ[104] deteriorates from 0.20 to 0.26 kcal mol$^{-1}$. However, we previously found the revised version of the latter, revDSD-PBEP86-D3BJ,[95] to be an improvement over DSD-PBEP86-D3BJ across the board, and this is also seen here for the new S66x8 reference, RMSD = 0.19 kcal mol$^{-1}$. Using the more up-to-date D4[105] dispersion correction, this latter RMSD drops to 0.17 kcal mol$^{-1}$ for revDSD-PBE86-D4. With RMSD of 0.10 kcal mol$^{-1}$, dRPA75[106] with a custom-fitted[42] D3BJ correction was the best performer in ref. 42 and

**Table 6** Root-mean-square deviations (RMSDs, kcal mol$^{-1}$) of different localized orbital coupled cluster methods with respect to the canonical DF-CCSD(T)/haVTZ level interaction energies of S66x8. Raw, CP, and half-CP represents the counterpoise uncorrected, full-, and half-CP corrected results. ΔCP represents the size of the counterpoise correction; ΔΔCP is the deviation between ΔCP with this particular localized method and with canonical CCSD(T)

| Methods | Coefficient (A) | | Total S66x8 | Hydrogen bonds | π stack | London Dispersion | Mixed Influence |
|---|---|---|---|---|---|---|---|
| | | | | | RMSD(kcal/mol) | | |
| PNO-CCSD(T)/default | | raw | 0.504 | 0.325 | 0.883 | 0.533 | 0.378 |
| | | CP | 0.136 | 0.112 | 0.229 | 0.120 | 0.107 |
| | | half | 0.319 | 0.218 | 0.556 | 0.326 | 0.242 |
| | | ΔΔCP | 0.371 | 0.215 | 0.655 | 0.415 | 0.273 |
| PNO-CCSD(T)/Tight | | raw | 0.232 | 0.138 | 0.344 | 0.300 | 0.193 |
| | | CP | 0.065 | 0.044 | 0.111 | 0.064 | 0.053 |
| | | half | 0.147 | 0.091 | 0.226 | 0.182 | 0.122 |
| | | ΔΔCP | 0.171 | 0.096 | 0.238 | 0.238 | 0.143 |
| Tight + A(Tight−Default) | 0.605 | raw | 0.143 | 0.063 | 0.237 | 0.187 | 0.111 |
| | 0.781 | CP | 0.028 | 0.020 | 0.048 | 0.026 | 0.023 |
| | 0.653 | half | 0.082 | 0.035 | 0.142 | 0.104 | 0.063 |
| | 0.516 | ΔΔCP | 0.123 | 0.059 | 0.188 | 0.168 | 0.098 |
| DLPNO-CCSD(T$_1$)/Tight | | raw | 0.134 | 0.146 | 0.195 | 0.115 | 0.086 |
| | | CP | 0.123 | 0.135 | 0.190 | 0.088 | 0.074 |
| | | half | 0.128 | 0.140 | 0.192 | 0.100 | 0.080 |
| | | ΔΔCP | 0.023 | 0.018 | 0.014 | 0.036 | 0.022 |
| DLPNO-CCSD(T$_1$)/Tight T$_{CutPNO}$=1.0×10$^{-6}$ | | raw | 0.279 | 0.282 | 0.410 | 0.261 | 0.190 |
| | | CP | 0.268 | 0.279 | 0.407 | 0.231 | 0.176 |
| | | half | 0.273 | 0.280 | 0.408 | 0.246 | 0.183 |
| | | ΔΔCP | 0.022 | 0.016 | 0.017 | 0.035 | 0.020 |
| DLPNO-CCSD(T$_1$)/VeryTight | | raw | 0.106 | 0.087 | 0.161 | 0.119 | 0.080 |
| | | CP | 0.081 | 0.070 | 0.137 | 0.079 | 0.053 |
| | | half | 0.093 | 0.078 | 0.149 | 0.099 | 0.066 |
| | | ΔΔCP | 0.029 | 0.018 | 0.028 | 0.043 | 0.029 |
| DLPNO-CCSD(T$_1$)/Tight CPS or T$_{CutPNO}$=1.0E−{6,7} | | raw | 0.073 | 0.081 | 0.106 | 0.059 | 0.049 |
| | | CP | 0.068 | 0.068 | 0.103 | 0.054 | 0.051 |
| | | half | 0.069 | 0.074 | 0.104 | 0.053 | 0.049 |
| | | ΔΔCP | 0.025 | 0.021 | 0.016 | 0.038 | 0.023 |
| LNO-CCSD(T)/Normal | | raw | 0.179 | 0.167 | 0.328 | 0.131 | 0.093 |
| | | CP | 0.225 | 0.138 | 0.375 | 0.262 | 0.175 |
| | | half | 0.196 | 0.150 | 0.351 | 0.184 | 0.132 |
| | | ΔΔCP | 0.111 | 0.064 | 0.068 | 0.193 | 0.096 |
| LNO-CCSD(T)/Tight | | raw | 0.068 | 0.088 | 0.091 | 0.043 | 0.031 |
| | | CP | 0.102 | 0.061 | 0.132 | 0.154 | 0.076 |
| | | half | 0.076 | 0.072 | 0.111 | 0.086 | 0.050 |
| | | ΔΔCP | 0.081 | 0.048 | 0.051 | 0.147 | 0.061 |
| LNO-CCSD(T)/vTight | | raw | 0.029 | 0.034 | 0.033 | 0.031 | 0.019 |
| | | CP | 0.051 | 0.016 | 0.065 | 0.084 | 0.039 |
| | | half | 0.029 | 0.018 | 0.046 | 0.035 | 0.023 |
| | | ΔΔCP | 0.060 | 0.039 | 0.044 | 0.107 | 0.042 |
| LNO-CCSD(T)/vvTight | | raw | 0.015 | 0.010 | 0.020 | 0.021 | 0.012 |
| | | CP | 0.056 | 0.035 | 0.062 | 0.090 | 0.041 |
| | | half | 0.027 | 0.017 | 0.039 | 0.037 | 0.021 |
| | | ΔΔCP | 0.061 | 0.040 | 0.048 | 0.108 | 0.044 |
| DF-CCSD(T) | | ΔCP | 0.722 | 0.585 | 1.165 | 0.688 | 0.581 |

continues to be so here (0.12 kcal mol$^{-1}$). [As explained in ref. 42, the D3BJ correction has a small coefficient and is similar to that found for CCSD: in effect, it compensates for the missing dispersion terms from (T), which start at fourth order in symmetry-adapted perturbation theory (SAPT)[107]]

As a parenthetical remark, the combinatorially optimized, range-separated hybrid, ωB97M-V[96] and double hybrid, ωB97M(2)[97] functionals were not yet available to us when ref. 42 was published; we find here 0.15 and 0.14 kcal mol$^{-1}$ root-mean-square deviations relative to the new reference, respectively. It means ωB97M-V outperforms all other rung 4 functionals, plus all rung 5 functionals considered in ref. 42 other than dRPA75-D3BJ and ωB97M(2).

DFT(SAPT),[108] *i.e.*, SAPT using DFT orbitals, has recently gained some currency as a relatively low-cost/high-accuracy approach for noncovalent interactions. Heßelmann[109] applied such approaches to S66x8 using asymptotically corrected PBE0ac orbitals and three different response kernels, namely ALDA (adiabatic local density approximation), TDEXX (time-dependent exact exchange), and ATDEXX (adiabatic TDEXX). The RMSD between Table 5 of his ESI† and the present "sterling silver" reference data is 0.224 kcal mol$^{-1}$ using ALDA, but this drops to 0.150 kcal mol$^{-1}$ for TDEXX and slightly further to 0.136 kcal mol$^{-1}$ for ATDEXX. Most of the improvement results from the π-stacked subset (see ESI† for details); suffice to say that both TDEXX and ATDEXX are competitive with the best DFT functionals considered here, and not greatly inferior to previous wavefunction calculation sets.

## IV. Conclusions

We have successfully calculated the "sterling silver" standard noncovalent interaction energies of the entire S66x8 set. Analyzing the RMS errors of three other S66x8 reference dissociation energies available in the literature, we can safely conclude that the "sterling silver" reference energies are markedly better than Hobza's original ones[37] as well as the earlier revised version proposed by our group,[42] but only marginally better than the "bronze"[43] level dissociation energies obtained at much lower cost.

Additionally, examining the RMS deviations of a variety of pure and composite localized coupled cluster methods for the new S66x8 reference, we can conclude the following:

(i) With half-CP, LNO-CCSD(T, vTight)/haV{T,Q}Z is among the top performers of all the pure and composite LNO-CCSD(T) methods tested. Although none of the low-cost cLNO methods are as good as the expensive LNO-CCSD(T,vTight)/haV{T,Q}Z half-CP, the Tight{T,Q} + 0.95[vTight − Tight]/T half-CP would be a viable alternative if someone is restricted to limited computational resources. Especially for London dispersion, the low-cost Tight{T,Q} + 0.78[vvTight − Tight]/T half-CP offers accuracy comparable to the more expensive LNO-CCSD(T, vTight)/haV{T,Q}Z half-CP or vTight{T,Q} + 0.31[vvTight − vTight]/T half-CP. For the intramolecular interactions, where counterpoise correction is not practical, we can safely recommend the composite method, vTight{T,Q} + 0.72[vvTight − vTight]/T.

(ii) Even with the Default threshold, PNO-LCCSD(T)/ haV{T,Q}Z full-CP performs remarkably sound—which gets even better with a tighter threshold. This remarkable result of the PNO-LCCSD(T, Tight)/haV{T,Q}Z full-CP could be due to the benefits from fortuitous error compensation. Using a composite scheme does not have any added advantage over the pure PNO-LCCSD(T) methods. However, for the intramolecular interactions, the low-cost cPNO method, Default{Q,5} + 0.15[Tight − Default]/T still has the edge over the pure methods due to its superiority for the π stack complexes.

(iii) Even with CPS and CBS extrapolation, the pure DLPNO-CCSD(T)] methods are not up to the mark for S66x8 noncovalent interactions. An RMS deviation of 0.059 kcal mol⁻¹ is the best accuracy we can achieve by employing a composite scheme, which is well behind the best picks among the PNO or LNO-based composite methods.

For the situations where any kind of CP correction is impractical, we recommend the composite cPNO and cLNO methods, Default{Q,5} + 0.15[Tight − Default]/T and Tight{T,Q} + 0.72[vvTight − vTight]/T, respectively.

Among more economical methods, the highest accuracies are seen for dRPA75-D3BJ, ωB97M-V, ωB97M(2), revDSD-PBEP86-D4, and DFT(SAPT) with a TDEXX or ATDEXX kernel.

## Note added in revision

While the present paper was being revised in response to peer reviewer comments, a study by Förster[110] appeared that compares the performance of RPA, second-order screened exchange[111] (SOSEX), and second-order statically screened exchange[112,113] for a variety of chemical problems, including S66x8. Extracting the RPA, RPA + SOSEX($W$, $\nu_c$), and RPA + SOSEX($W(0)$, $W(0)$) level interaction energies from the ESI of ref. 110, we have evaluated the RMS deviations relative to the "sterling silver" reference data (see the ESI†). With full-CP correction and CBS extrapolation, RPA, RPA + SOSEX($W$, $\nu_c$), and RPA + SOSEX($W(0)$, $W(0)$) has RMSD of 0.44, 0.35, 0.40 kcal mol⁻¹, respectively—note that none of these include separate dispersion corrections, unlike the better-performing DFT options in the present paper.

## Conflicts of interest

There authors declare no conflicts of interest.

## Acknowledgements

This research was funded by the Israel Science Foundation (grant 1969/20), the Minerva Foundation (grant 20/05), as well as by a research grant from the Artificial Intelligence and Smart Materials Research Fund, in Memory of Dr Uriel Arnon, Israel. AK gratefully acknowledges an Australian Research Council (ARC) Future Fellowship (Project No. FT170100373). GS acknowledges a doctoral fellowship from the Feinberg Graduate School (WIS). The work of E. S. on this scientific paper was supported by the Onassis Foundation—Scholarship ID: FZP 052-2/2021-2022. AK acknowledges the system administration support provided by the Faculty of Science at the University of Western Australia to the Linux cluster of the Karton group. We thank Dr G. Bistoni (University of Perugia, Italy) for a clarification concerning the complete pair natural orbital space (CPS) extrapolation exponent in ref. 92.